\documentclass[]{aa}

\usepackage{psfig}

\begin{document}

\title{Photometric structure of polar-ring galaxies}
\author{V. P. Reshetnikov\inst{1,}\inst{2}}

\offprints{resh@astro.spbu.ru }

\institute{Astronomical Institute of St.\,Petersburg State 
University, 198504 St.\,Petersburg, Russia 
\and 
Isaac Newton Institute of Chile, St.\,Petersburg Branch}

\date{Received 3 June 2003 / Accepted  21 August 2003}

\titlerunning{Polar-ring galaxies}

\authorrunning{Reshetnikov}

\abstract{
The results of $B$, $V$, $R$ surface photometry of three polar-ring
galaxies (PRGs) -- A\,0017+2212, UGC\,1198, UGC\,4385 --
are presented. 
The data were acquired at the 6-m telescope
of the Special Astrophysical Observatory of the Russian Academy of Sciences.
It was shown that all three galaxies are peculiar late-type spirals
in the state of ongoing interaction or merging. \\
We discuss available photometric properties of the PRGs with
spiral hosts and 
consider the Tully-Fisher relation for different types of PRGs. 
In agreement with Iodice et al. (2003), we have shown that true
PRGs demonstrate $\sim$1/3 larger maximum rotation velocities than
spiral galaxies of the same luminosity. Peculiar objects with
forming polar structures satisfy, on average, the Tully-Fisher
relation for disk galaxies but with large scatter. 
}

\maketitle
\section{Introduction}

Polar-ring galaxies (PRGs) are peculiar objects composed of a central
component surrounded by an outer nearly perpendicular ring or
disk, made up of gas, stars and dust (Whitmore et al. 1990, Polar
Ring Catalog=PRC). 
It is generally agreed 
that the formation of the polar 
rings is the result of some
 kind of ``secondary event'' -- external 
accretion or a merger.
 Recent numerical simulations have confirmed
that both scenarios are robust, and result in realistic polar
structures (Bekki 1997, 1998; Reshetnikov \& Sotnikova 1997;
Bournaud \& Combes 2003).

PRGs are ideal laboratories to study the shape of their dark matter
haloes, and gain insight into the galaxy formation through late
infall, because of the high inclination of their outer structures
and the evidence of very recent or occuring gas infall.
PRC lists more than 100 PRGs and candidates for PRGs. Unfortunately,
only a small fraction of them are investigated. The last years
have seen renewed interest in PRG studies, triggered
by the wonderful data acquired during the HST heritage program
for the polar-ring galaxy NGC\,4650A (Kinney et al. 1999).

In this article we present the results of photometric observations
of three possible candidates for PRGs from the PRC. One of the galaxies --
UGC\,4385 -- is a true, kinematically-confirmed polar-ring galaxy
with a spiral host (Reshetnikov \& Combes 1994). Two other galaxies
(A\,0017+2212 and UGC\,1198) are almost uninvestigated peculiar
objects. In the last section of the paper we discuss the available
data on the photometric structure of PRGs and re-examine the 
Tully-Fisher relation for various groups of PRGs. 

\section{Observations and reductions}

The photometric observations were carried out in 1992 and 1996
in the prime focus of the 6-m telescope of the Special Astrophysical
Observatory of the Russian Academy of Sciences, using a CCD detector
with 520$\times$580 pixels, each 18$\times$24 $\mu$m 
(0\farcs15 $\times$ 0\farcs20) (Borisenko et al. 1991).
The data were acquired with standard Johnson $B$, $V$ and Cousins
$R$ filters. Photometric calibration was accomplished using repeated
observations of standard stars from the Landolt (1983) and Smith et al.
(1991) lists. The seeing during the observations was 1\farcs8
(A\,0017+2212, UGC\,4385) and 1\farcs0 for UGC\,1198. A log of
observations is given in Table 1 (extinction corrected sky
brightness (in mag arcsec$^{-2}$) in each frame is presented in
the last column of the table). Reduction of the CCD data has been 
performed in the standard manner using the ESO-MIDAS\footnote{MIDAS is 
developed and maintained by the European Southern Observatory.}
package.

\begin{table}
\caption{ Observations }
\begin{center}
\begin{tabular}{|c|c|c|c|c|c|}
\hline
Object &  Data & Band-  & Exp &  $Z$    & Sky  \\
       &       & pass   & (sec)& ($^{\rm o}$)& mag. \\
\hline                   
A\,0017+2212     & Nov 1992 & R   & 300 & 31 & 19.6\\
 (PRC C-2)     &           & R   & 300 & 30 &19.6 \\
	       &           & V   & 720 & 29 & 20.0\\
	       &           & B   & 900 & 27 & 20.5\\
	       &           & B   & 900 & 25 & 20.5\\
               &           &     &     &       & \\
UGC\,1198      & Nov 1992 & R  & 300 & 42 & 20.4 \\
 (PRC C-12)     &           & V  & 600 & 42 & 20.9 \\
		&           & B  & 900 & 42 & 21.8 \\
		&           & B  & 900 & 42 & 21.8\\
		&           &    &     &       & \\
UGC\,4385       & May 1996&  R  & 300 & 33 & 20.6 \\	       
 (PRC C-27)     &          &  V  & 600 & 32 & 21.2 \\ 
                &          &  B  & 900 & 31 & 22.1 \\
\hline
\end{tabular}
\end{center}
\end{table}

The total magnitudes of galaxies found by us from the multiaperture
photometry (see Table 2) are in good agreement with the 
NED\footnote{NASA/IPAC Extragalactic Database.} data
($B_T=16.75$, 14.80$\pm$0.16, 14.51$\pm$0.18 for A\,0017+2212, UGC\,1198,
UGC\,4385 correspondingly). The mean difference between our and
NED total magnitudes is --0.07$\pm$0.15($\sigma$).

\section{Results and discussion}

\subsection{A\,0017+2212 (PRC C-2)}

\subsubsection{General structure}

The $B$-band image is displayed in Fig. 1; Fig. 2 shows $V$-band
contour map. The galaxy demonstrates a multicomponent structure. 
The central part of A\,0017+2212 is elongated along the position
angle close to 0$^{\rm o}$ (north--south direction). The central
body is surrounded by a narrow inclined (by about 30$^{\rm o}$)
ring-like structure with angular diameter $\approx$20$''$ and
apparent axial ratio $b/a=0.4$. Two bright knots are located
at the edges of the ring (Fig. 1, Fig. 4). The central body and the
ring are embedded in an asymmetric extended envelope. A small edge-on
galaxy is located 45$''$ SW of the PRC object.

\begin{figure}
\centerline{\psfig{file=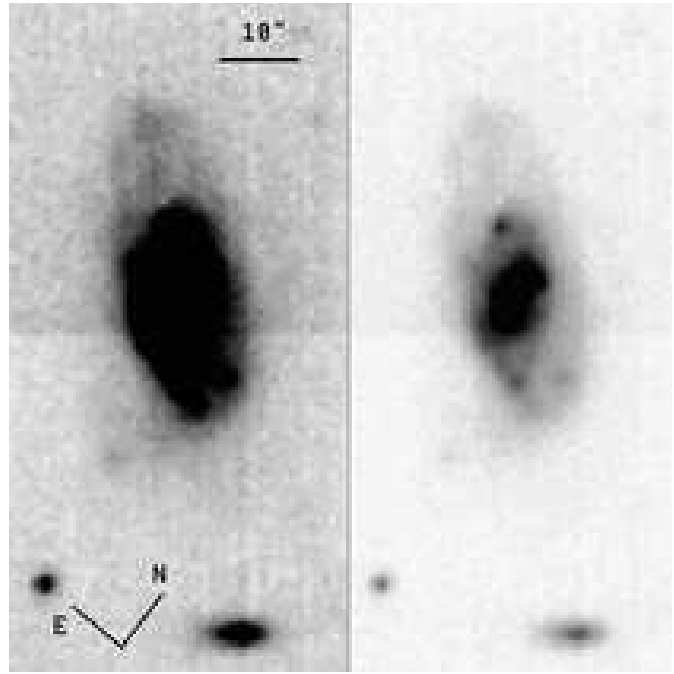,width=8.8cm,angle=0,clip=}}
\caption{$B$-band image of A\,0017+2212 displayed at two different
contrasts. Each image size is 42\arcsec $\times$ 85\arcsec.
}
\end{figure}

\begin{figure}
\centerline{\psfig{file=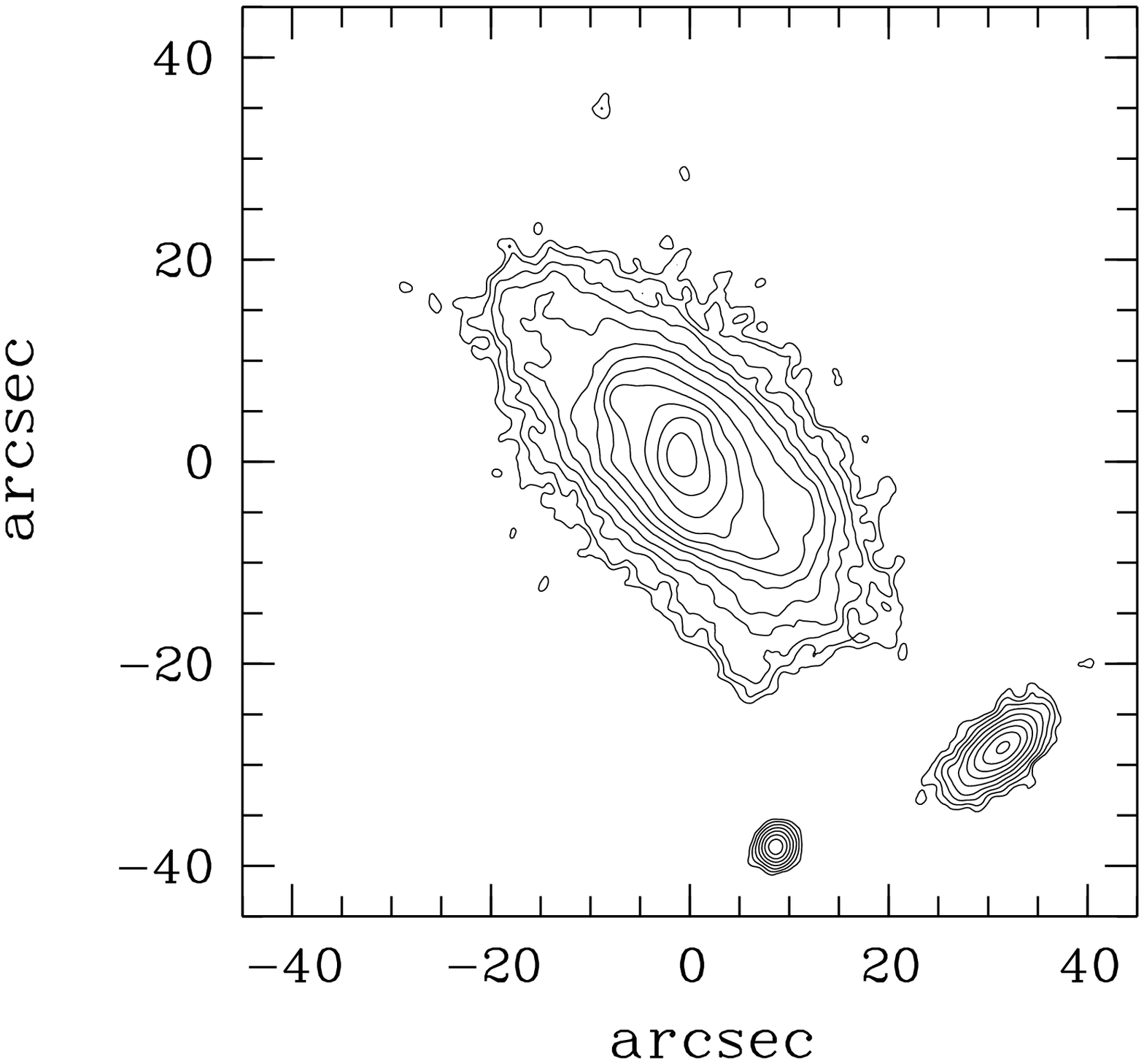,width=8.8cm,angle=-90,clip=}}
\caption{Isophotal contour image of A\,0017+2212 in the $V$ passband.
The faintest contour is 25.7 mag arcsec$^{-2}$, isophotes step --
0\fm44. North is at the top, east to the left.
}
\end{figure}

Figures 3a,c show the surface brightness profiles  
along the apparent major axis of the ring (P.A.=43$^{\rm o}$) and
along the major axis of the central part of the galaxy
(P.A.=13$^{\rm o}$).
In the major axis profile the ring is seen as two symmetrical
maxima at $r \approx \pm 10^{''}$. Excluding the regions of the
ring projection, the general surface brightness distribution 
may be approximated by that of an exponential disk. In the $R$
passband the disk characteristics are: the central surface 
brightness $\mu_{R,0}(0)$=19.6\footnote{Subscript ``0'' means
value corrected for Galactic absorption according to 
Schlegel et al. (1998)}, exponential scalelength
$h$=3$''$ at P.A.=13$^{\rm o}$
(or $h$=4$''$ at P.A.=43$^{\rm o}$). In the $B$-band the extrapolated
central surface brightness of the disk is $\mu_{B,0}(0)$=20.6.
This value is about 1$^m$ brighter than that of the disks of
normal bright spirals (Freeman 1970). Such bright galactic disks 
are usual for interacting systems (Reshetnikov et al. 1993).

Figures 3b,d display the observed color indices along two cuts.
Both profiles show strong color gradients: the central regions
of the galaxy are relatively red ($(B-V)_0\approx+0.7$), while
the outer ones are blue ($(B-V)_0\approx+0.4$). 

\begin{figure}
\centerline{\psfig{file=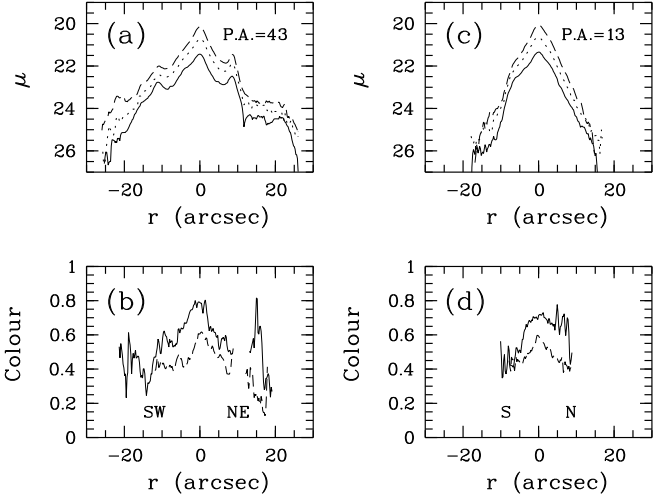,width=8.8cm,angle=-90,clip=}}
\caption{Photometric profiles for A\,0017+2212: {\bf a)}, {\bf b)}
along the apparent major axis (through two bright knots in the ring);
{\bf c)}, {\bf d)} along the major axis of the central component.
Solid line in {\bf a)} and {\bf c)} represent the distributions in
the $B$ passband, dotted lines in $V$, and dashed ones in the $R$.
Solid lines in {\bf b)} and {\bf d)} show the distribution of the
$B-V$ color, dashed in the $V-R$ color.}
\end{figure}

From optical colors (see Buta et al. 1994; Buta \& Williams 1995)
and the exponential surface brightness distribution we can preliminary 
classify the galaxy as an Sc/Scd spiral. 

\subsubsection{Ring}

The suspected ring looks very narrow -- with the ratio of
radial extent to mean radius about 20\% (Figs. 1,4). The total apparent
magnitude of the ring is $B = 19^m$, or the ring contribution to
the total luminosity of A\,0017+2212 is $\approx$1/9.
The mean optical colors of the ring are close to that for whole
galaxy: $(B-V)_0=+0.48$, $(V-R)_0=+0.38$. In general, the NE arc of
the ring is redder ($(B-V)_0\approx+0.55$) than the SW part
($(B-V)_0\approx+0.4$).

Several bright knots are evident in the ring (star clusters, giant
HII regions?). The two brightest knots (with $B\approx21.5-22$ through
2$''$ circle aperture) are 
situated near the edges of the ring (Figs. 1,4). Optical colors of
two brightest knots are somewhat bluer in comparison with
nearby parts of the ring: $(B-V)_0\approx+0.5$ for the NE knot,
and $(B-V)_0\approx+0.3$ for the SW (Fig. 4). 

\begin{figure}
\centerline{\psfig{file=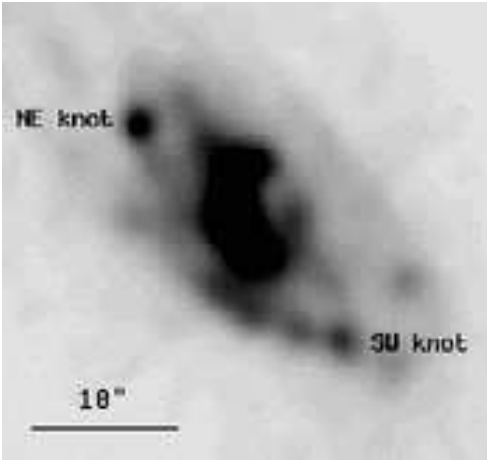,width=6.5cm,clip=}}
\caption{$B$-band image of A\,0017+2212 after Lucy--Richardson
restoration (ten iterations). North is at the top, east to the left.}
\end{figure}

\subsubsection{Companion}

The SW companion is faint (total apparent magnitude
$B_T=20.0\pm0.1$) and almost edge-on galaxy (Figs. 1,3,5).
The galaxy coordinates are $\alpha=00^h19^m50.0^s$,
$\delta=+22^{\rm o}28'14''$ (J2000) within 1$''$ of error. 
Therefore, we have denoted this object Anon\,J001950.0+222814.
Integrated optical colors of the galaxy -- $(B-V)_0=+0.63\pm0.10$,
$(V-R)_0=+0.48\pm0.05$ -- are typical for an Sab/Sb spiral
(Buta et al. 1994; Buta \& Williams 1995).

The radial surface brightness distribution of Anon\,J001950.0+222814
can be approximated by an exponential law with $\mu_{R,0}(0)$=21.1
($\mu_{B,0}(0)$=22.1) and $h=2\farcs25\pm0\farcs5$. The radial
structure of the disk is strongly asymmetric -- $h=2\farcs75$ for
the SE part of the disk, $h=1\farcs75$ for the NW side. 

From the analysis of the vertical structure of the galaxy, we
found evidence of a flaring stellar disk. A minor axis photometric
cut in the $R$ passband gives an exponential scaleheight value of
$h_z$=0\farcs9. Cuts at distances $r$=2\farcs2--2\farcs9 from the nucleus 
give larger values: $h_z$=1\farcs1. The observed $h/h_z$ ratio
is rather small: 2--2.5. PSF correction increases this ratio
to 3--4. Therefore, the stellar disk of Anon\,J001950.0+222814 is 
unusually thick in comparison with typical Sab/Sb spiral galaxies 
(e.g. de Grijs 1998; Schwarzkopf \& Dettmar 2000). Such thick
stellar disks are typical of interacting galaxies (Reshetnikov \&
Combes 1997; Schwarzkopf \& Dettmar 2000).

The southeast part of Anon\,J001950.0+222814 shows a faint 
tidal extention pointing to A\,0017+2212 (Fig. 5). 

\begin{figure}
\centerline{\psfig{file=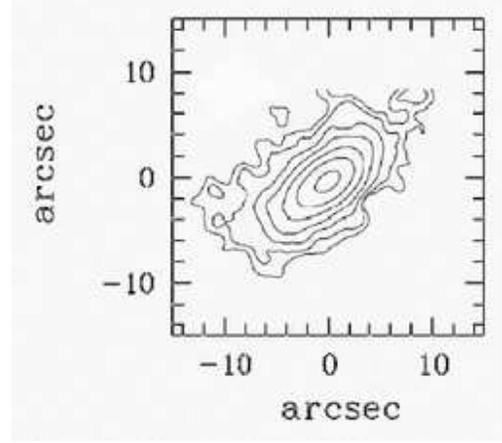,width=6.5cm,clip=}}
\caption{Isophotal contour image of Anon\,J001950.0+222814
in the $V$ filter. The fainest contour is $\sim$26.5 mag arcsec$^{-2}$,
isophotes step -- 0\fm75. North is at the top, east to the left.}
\end{figure}

\begin{table*}
\caption{ General characteristics of PRGs }
\begin{center}
\begin{tabular}{|l|ccc|l|}
\hline
Parameter &  A\,0017+2212  & UGC\,1198 &  UGC\,4385 & Ref. \\
\hline                   
                   &            &        &          &   \\
Morphological type &   Sc/Scd   &  Sc:   &   Im:    &    \\
Heliocentric velocity (km/s) &   & 1172\,$\pm$\,8  & 1969\,$\pm$\,10 & NED  \\
Distance (Mpc) (H$_0$=70 km/s Mpc$^{-1}$) &        &  19.2 &  26.6  &     \\
Major axis, $D_{25}$ ($\mu_B=25$) & 46\arcsec & 54\arcsec (5.0 kpc) & 70\arcsec: (9.0 kpc) &      \\
Axial ratio, $b/a$ ($\mu_B=25$)  & 0.52 & 0.79 & 0.9: & \\
                                 &   &  &  & \\
Total apparent magnitudes        &   &  &  & \\
and colors:                      &   &  &  & \\
$B_T$           & 16.60\,$\pm$\,0.10 & 14.90\,$\pm$\,0.10 & 14.35\,$\pm$\,0.10 &\\
$(B-V)_T$       & +0.60\,$\pm$\,0.03 & +0.70\,$\pm$\,0.03 & +0.46\,$\pm$\,0.04 &\\
$(V-R)_T$       & +0.43\,$\pm$\,0.03 & +0.45\,$\pm$\,0.03 & +0.34\,$\pm$\,0.04 &\\
$(J-H)_{2MASS}$ & +0.64\,$\pm$\,0.09 &          & +0.52\,$\pm$\,0.06   & [1]\\
$(H-K)_{2MASS}$ & +0.22\,$\pm$\,0.13 &          & +0.20\,$\pm$\,0.09  & [1]\\
                                 &   &  &  &  \\
Galactic absorption ($B$-band) & 0.34  & 0.51    &  0.19  & NED\\
Absolute magnitude, $M_{B,0}$  &   &    --17.0   & --18.0     & \\
$(B-V)_0$                    & +0.52 & +0.58     &  +0.41  & \\
$(V-R)_0$                    & +0.38 & +0.37     &  +0.32  & \\
                                 &   &  &  &  \\
Exponential disk:                &   &  &  &  \\
\hspace*{3cm}$\mu_{R,0}(0)$          & 19.6 & 19.1 & 19.3: &  \\
\hspace*{3cm}$h(R)$                 & 4\farcs0 & 6\farcs4 (0.59 kpc) & 7\farcs3: (0.94 kpc) & \\
                                 &   &  &  &  \\
Ring structure:                &   &  &  &  \\
$(B-V)_0$                      & +0.48  &      & +0.35 &  \\
$(V-R)_0$                      & +0.38  &      & +0.25 &  \\
Ring-to-galaxy ratio ($B$-band)& 0.1:   &  & 0.1: &  \\
                                 &   &  &  &  \\
Far-infrared luminosity ($L_{\odot}$) & & 1.4\,10$^9$ & 4.7\,10$^8$  & NED  \\				 
M(HI) (M$_{\odot}$)          &   & (1.7\,$\pm$\,0.9)\,10$^8$ & 1.1\,10$^9$ & [2] \\
M(H$_2$) (M$_{\odot}$)       &   & (0.8\,$\pm$\,0.2)\,10$^8$ &  &  [3]\\
M(HI)/$L_B$ (M$_{\odot}/L_{B,\odot}$)       &   & 0.17 & 0.45 & \\
W$_{20}$ (km/s)                 &   & 156\,$\pm$\,13 & 196 & [2]\\
\hline
\end{tabular}
\end{center}
[1] -- Skrutskie et al. (1997), [2] -- van Driel et al. (2000),
[3] -- Galletta et al. (1997)
\end{table*}

\subsubsection{General results on A\,0017+2212}

In general photometric characteristics (surface brightness
distribution, integrated colors (Table 2), color gradients) A\,0017+2212
looks similar to late-type spiral galaxies. The main peculiarity
of the galaxy is a narrow ring-like structure surrounding
central part of the galaxy. This ring is inclined some 30$^{\rm o}$
with respect to the inner region of the galaxy (Figs. 1, 2, 4).
The galaxy is embedded in a faint and knotty outer envelope.
The major axis of the envelope coincides with the major axis of
the ring. 

The relatively faint companion (Anon\,J001950.0+222814) lies 45$''$ SW of 
A\,0017+2212. Important features of the companion are:
\begin{itemize}
\item asymmetric radial distribution of the surface brightness;
\item thick stellar disk;
\item tidal tail pointing to A\,0017+2212;
\item the companion is located approximately along
the major axis of the ring and of the envelope of A\,0017+2212. 
\end{itemize}
All the above features suggest an ongoing interaction of the companion with
the main galaxy. One can speculate that we observe the very rare and
brief stage of inclined ring formation due to tidal stripping
of outer parts of the donor galaxy (Anon\,J001950.0+222814)
(Reshetnikov \& Sotnikova 1997; Bournaud \& Combes 2003). 
Therefore, A\,0017+2212 can be an example of a strongly inclined ring
around a spiral host. Other examples of such objects are NGC\,660
(van Driel et al. 1995) and ESO\,235-G58 (Iodice 2001). 

In an alternative scenario, the observed morphology of A\,0017+2212
is the result of a high-velocity collision between the companion
galaxy (Anon\,J001950.0+222814) and a larger disk system. Therefore,
A\,0017+2212 can be a ``Cartwheel-type'' collisional ring
galaxy (e.g. Appleton \& Struck-Marcell 1996). It is
also possible that A\,0017+2212 is a peculiar ringed spiral with
a bar (see examples in Buta \& Combes 1996). In this case the
inner structure elongated at P.A.$\approx 0^{\rm o}$ can be considered
as a bar.

Unfortunately, A\,0017+2212 has no published redshift and we cannot
describe the galaxy in absolute units. For a crude scaling, we
will assume that exponential scalelength of A\,0017+2212 is close
to the scale of NGC\,660 -- the galaxy with inclined ring
($h=1.4$ kpc for H$_0$=70 km/s Mpc$^{-1}$ according to van Driel. 1995). 
This leads to an absolute magnitude $M_{B,0}$=--18.6 (close to
that for NGC\,660) and the ring radius of about 5 kpc. The projected
distance to the companion thus 21 kpc. 

Table 2 summarizes the main characteristics of A\,0017+2212 and
two other PRGs, both found in this work and collected from the 
literature.

\subsection{UGC\,1198}

\subsubsection{Photometric structure}

UGC\,1198 demonstrates almost elliptical outer isophotes
(Figs. 6, 7). A faint ($B_T=22.6$) elliptical object (background galaxy?) 
is located 17$\arcsec$ NW of the galaxy center. UGC\,1198
shows disk-like enhancement along the minor axis and two linear
structures elongated approximately
along the main body of the galaxy (Fig. 6). A faint
ring-like feature is seen to the east of the center. 
The isophotal map (Fig. 7) shows large-scale twisting: the position angle
of the outer isophotes (P.A.$\approx115^{\rm o}$) is shifted by about
22$^{\rm o}$ relative to those of the central ones 
(P.A.$\approx93^{\rm o}$).

\begin{figure}
\centerline{\psfig{file=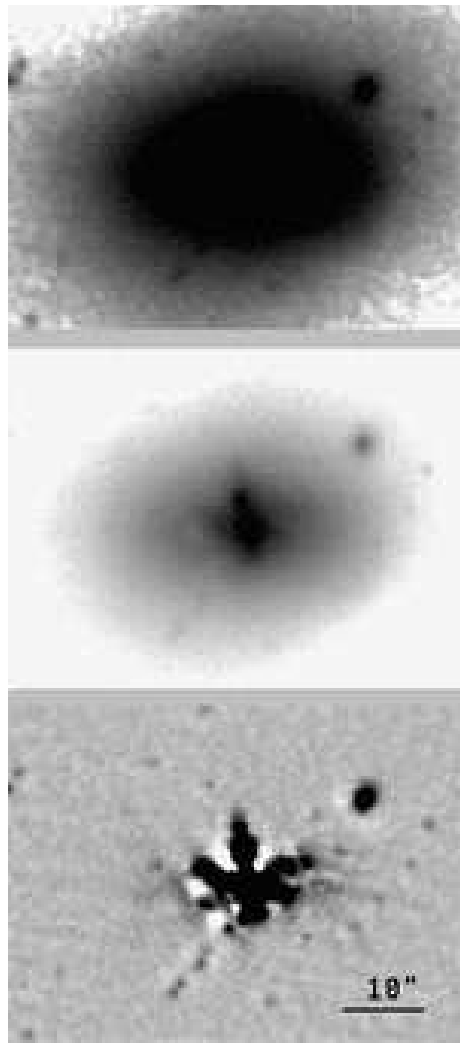,width=6.0cm,angle=0,clip=}}
\caption{$R$-band images of UGC\,1198. The first two images are
the same frame displayed at different contrasts. Bottom frame is
the residual image of the subtraction of the 4\farcs3$\times$4\farcs3
median-filtered frame from the original $R$ image. Each image size is 
57\arcsec $\times$ 42\arcsec. North is at the top, east to the left.
}
\end{figure}

\begin{figure}
\centerline{\psfig{file=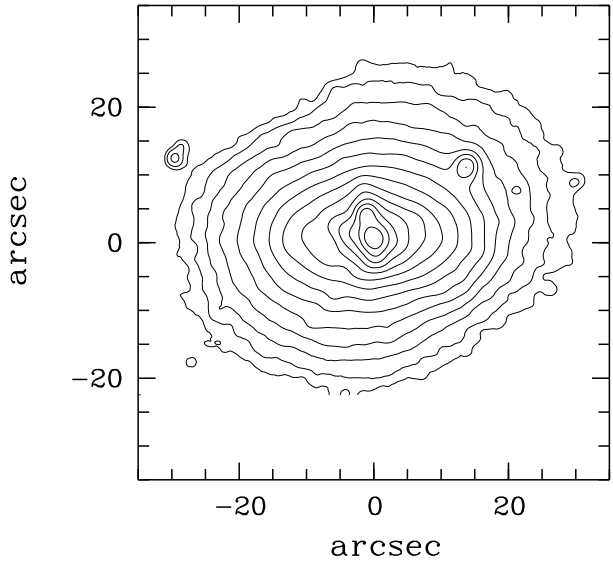,width=8.4cm,angle=-90,clip=}}
\caption{Isophotal contour image of UGC\,1198 in the $V$ passband.
The faintest contour is 25.0 mag arcsec$^{-2}$, isophotes step --
0\fm44. North is at the top, east to the left.
}
\end{figure}

Photometric profiles approximately along the major and minor axes
are shown in Fig. 8. Out of the central 5\arcsec--10\arcsec, the surface
brightness is nearly exponential with a central surface brightness
$\mu_{R,0}(0)$=19.1 and exponential scalelength $h=6\farcs4$.
In the $B$ passband $\mu_{B,0}(0)$=19.8 that is significantly
brighter than typical values for spiral disks. 
Observed color distributions (Fig. 8b,d) show several blue
patches crossing the central region of the galaxy. Another interesting
feature is the systematical reddening in the $B-V$ color from
the center to the outer parts. 

\begin{figure}
\centerline{\psfig{file=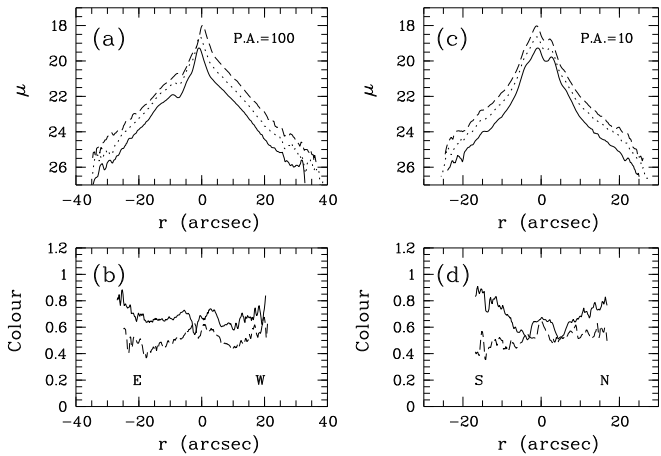,width=8.8cm,angle=-90,clip=}}
\caption{Photometric profiles for UGC\,1198: {\bf a)}, {\bf b)}
along the apparent major axis;
{\bf c)}, {\bf d)} along the major axis of inner elongated component.
Solid line in {\bf a)} and {\bf c)} represent the distributions in
the $B$ passband, dotted lines in $V$, and dashed ones in the $R$.
Solid lines in {\bf b)} and {\bf d)} show the distribution of the
$B-V$ color, dashed in the $V-R$ color.}
\end{figure}

Figure 9 gives the two-dimensional distribution of observed 
color index $B-V$ in the central region of UGC\,1198. A blue (white
in the figure) structure with $(B-V)_0=+0.38\pm0.05$ is elongated 
along the minor axis. The observed morphology of this feature suggests
that it can be an almost edge-on warped disk or ring. Blue colors 
give evidence of active star-formation in the disk. The blue 
structure is crossed by redder patches. Dark regions to the E and W
of the center in Fig. 9 show $(B-V)_0\approx+0.6$. The tangled
inner structure of UGC\,1198 makes it difficult to discuss it in
more detail.

\begin{figure}
\centerline{\psfig{file=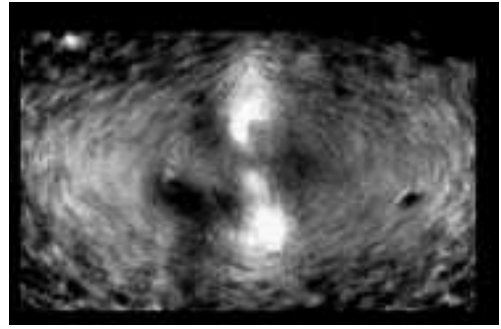,width=6.5cm,clip=}}
\caption{A gray-scale map of the $B-V$ colour index distribution
within central part (31\farcs6$\times$19\farcs5) of UGC\,1198.
The scale is such that $dark$ to $light$ is the sense $red$ to
$blue$. North is at the top and east to the left. }
\end{figure}

\subsubsection{Fine structures}

In order to emphasize small-scale internal structures of UGC\,1198,
we present in Fig. 6 (bottom frame) the residual image of the 
subtraction of a median-filtered image from the original $R$ frame. 
An inner disk at P.A.=10$^{\rm o}$ is clearly seen. Also, two features
elongated at P.A.$\approx$70$^{\rm o}$ and $\approx$110$^{\rm o}$
intersect the galaxy center. 

Some faint point-like sources are seen in the galaxy body (Fig. 6). 
The brightest sources form a jet-like structure at P.A.=139$^{\rm o}$. 
Aperture photometry (diameter of the aperture is 1\farcs6) of 14 clumps 
gives the following average characteristics: $B_0=25.8\pm0.4$ or
$M_{B,0}=-5.6$, $(B-V)_0=+0.7\pm0.3$, $(V-R)_0=+0.7\pm0.3$. 
One can propose that these sources are supergiant stars
or compact star clusters.

\subsubsection{Main results on UGC\,1198}

This small galaxy demonstrates an exponential surface brightness
distribution (Fig. 8) and optical colors typical for a Sbc/Scd
galaxy (Table 2). The detected relative quantities of HI and
H$_2$ (Table 2) also correspond to a late-type spiral galaxy 
(Young \& Knezek 1989). 

The far-infrared luminosity of UGC\,1198 estimated from the
flux densities at 60 $\mu$m and 100 $\mu$m (NED; Lonsdale et al. 
1989) is $L_{FIR}$=1.4$\times$10$^9$\,$L_{\odot}$. Therefore,
the far-infrared to blue luminosity ratio is 2.25, which is
much higher than for isolated spirals (0.4--0.6 -- de Jong et al. 1984;
Bushouse et al. 1988). Such a high value of the $L_{FIR}/L_B$ ratio
is usual for interacting galaxies and for forming PRGs 
(Bushouse et al. 1988; Reshetnikov \& Combes 1994). The 
enhanced value of $L_{FIR}/L_B$ gives evidence for active star 
formation in UGC\,1198. The star-formation is probably concentrated
in a gaseous disk which rotates with V$_{\rm max} \approx 70$ km/s
(Table 2) around the major axis of the galaxy. 
 
The complicated morphology of UGC\,1198 suggests that we observe
the result of a merger event. One can speculate that UGC\,1198
represents an actual collision between a small late-type galaxy and 
a small early-type galaxy moving on a nearly polar orbit relative to 
a pre-existing spiral. Tidal debris of the wrecking galaxy has created
a red outer envelope and the central part of it is still seen as the
object 17$\arcsec$ NW of the UGC\,1198 center (Fig. 6).

Final conclusions about the nature of this peculiar galaxy can be
obtained from additional high-resolution kinematical observations.

\subsection{UGC\,4385}

\subsubsection{General structure}

Optical images of the galaxy (see Fig. 10 and 11) show two distinct
morphological components: (1) a central body with a major axis position
angle of about 25$^{\rm o}$, and (2) a polar ring or disk with a
mean P.A. of $\approx$110$^{\rm o}$. The galaxy is surrounded by
a faint envelope whose major axis is aligned with the major axis
of the central body. The central body shows clear twisting of the
isophotes: from P.A. close to 0$^{\rm o}$ near the nucleus to
P.A.$\approx$25$^{\rm o}$ at $r \geq 30\arcsec$ (Fig. 11).

\begin{figure}
\centerline{\psfig{file=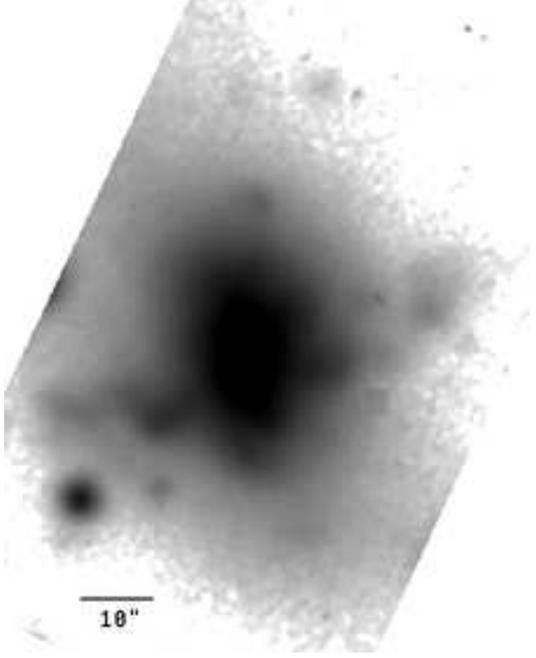,width=7cm,angle=0,clip=}}
\caption{$B$-band image of UGC\,4385. Image size is 
76\arcsec $\times$ 94\arcsec.
 North is at the top and east to the left.}
\end{figure}

\begin{figure}
\centerline{\psfig{file=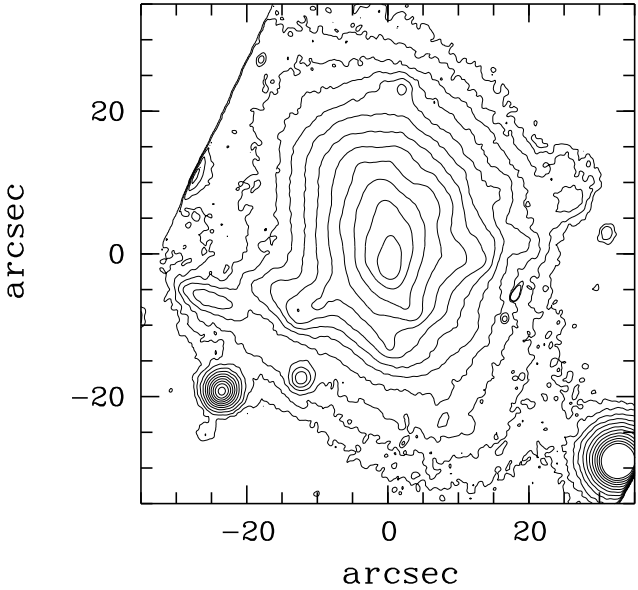,width=8.8cm,angle=-90,clip=}}
\caption{Isophotal contour image of UGC\,4385 in the $V$ passband.
The faintest contour is 25.0 mag arcsec$^{-2}$, isophotes step --
0\fm44. North is at the top, east to the left.
}
\end{figure}

The emission-line rotation curves of UGC\,4385 demonstrate that
the galaxy's central body rotates around the minor axis while
the ring rotates around the major axis (Reshetnikov \& Combes 1994).
Therefore, UGC\,4385 is one of the few true polar-ring galaxies 
with a spiral host (the other well-known example is NGC\,660 -- 
see van Driel et al. 1995). This conclusion is based on optical and
near-infrared colors, surface brightness distribution and HI content (see
discussion below and Table 2).

Fig. 12 shows the photometric profiles along the major and minor
axes of the galaxy. The central body shows an almost exponential
surface brightness distribution along the apparent major axis
at $r \leq 20\arcsec$ (Fig. 12a). At $r \geq 20\arcsec$ the
surface brightness slope becomes less steep, probably due to
the outer envelope contribution. The southern part of the major axis
profile shows some excess in comparison with the pure exponent
(Fig. 12a). The extinction-corrected extrapolated surface brightness
of the disk -- $\mu_{R,0}(0)$=19.3: ($\mu_{B,0}(0)$=19.9:) -- 
is brighter than in
typical spiral disks. External parts of the polar ring are clearly
seen as a local surface brightness maximum at $r \approx \pm25\arcsec$
(Fig. 12c). The major axis profile shows strong color gradient:
the central part of the galaxy is rather blue ($B-V \approx +0.35$),
while the outer parts are red ($B-V \approx +0.7$) (Fig. 12b).

UGC\,4385 is a faint source of far-infrared emission (Table 2).
The galaxy demonstrates a relatively small $L_{FIR}/L_B$ ratio (0.32)
which is typical for unperturbed spirals. 

The optical and near-infrared colors of UGC\,4385
as far as relative content of HI (see Table 2) are usual for 
a late-type disk galaxy.

\begin{figure}
\centerline{\psfig{file=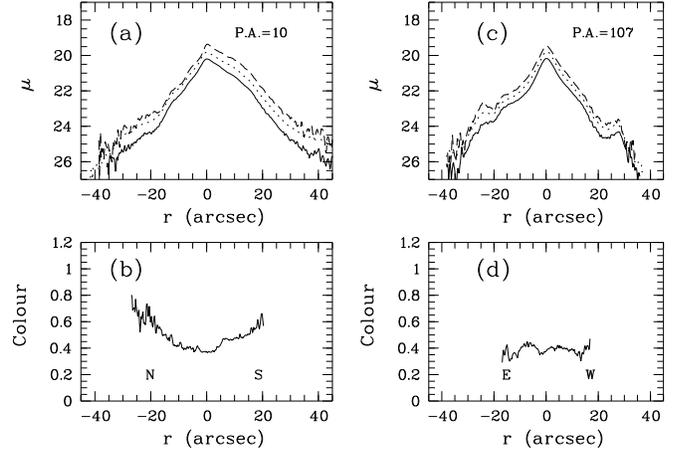,width=8.8cm,angle=-90,clip=}}
\caption{Photometric profiles for UGC\,4385: {\bf a)}, {\bf b)}
along the major axis of central body;
{\bf c)}, {\bf d)} along the major axis of the ring.
Solid line in {\bf a)} and {\bf c)} represent the distributions in
the $B$ passband, dotted lines in $V$, and dashed ones in the $R$.
Solid lines in {\bf b)} and {\bf d)} show the distribution of the
$B-V$ color.}
\end{figure}

\subsubsection{Ring}

The ring looks knotty, irregular and slightly inclined to the
line of sight. From the apparent axial ratio we esimate the ring's 
inclination as 80$^{\rm o}$-85$^{\rm o}$. Also, the ring 
protrudes to the south (Figs. 10, 11) The ring is quite
extended -- its standard isophotal radius 
$r(\mu_B=25)$ is 30$\arcsec$ or about 4 exponential scalelenghts of
the main body.

In Fig. 13 we present the residual image of the galaxy. 
In the main body we see a strongly warped asymmetric structure 
with axial ratio $b/a \leq 0.2$ (edge-on disk?). 
Numerous condensatons are located in the corrugated
ring and also are spreaded along the major axis of the central
body (Fig. 13). Typical $B$-band apparent magnitudes of the
condensations are about 22\fm5--23$^m$, which corresponds to an
absolute magnitude $M_B \approx$--9.1 -- --9.6. Most probably,
the condensations represent young star clusters which are
very numerous in interacting and merging galaxies (e.g. Schweizer
2002).

\begin{figure}
\centerline{\psfig{file=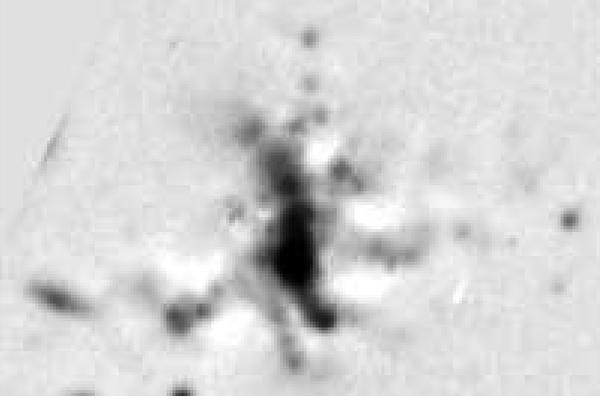,width=8.0cm,angle=0,clip=}}
\caption{Residual image of the subtraction of the 8\farcs4$\times$8\farcs4
median-filtered frame from the original $R$ image. Image size is 
69\arcsec $\times$ 46\arcsec. North is at the top, east to the left.
}
\end{figure}

Fig. 14 shows smoothed emission-line rotation curves of the galaxy
with the slit positions along the major axis of the main body
and along the polar ring. At $r \geq 14\arcsec$ ($\geq$1.8 kpc)
the observed polar and equatorial (along the galaxy's major axis)
velocities are almost equal. At the last point where measurements
along two ortoghonal directions are available ($r = 26\farcs2 =
3.6h$ = 3.4 kpc), the observed V$_{pol}$/V$_{eq}$ ratio is
1.04$\pm$0.22. After applying a small and rather uncertain correction
for non-edge-on orientation of the central body and the ring,
this ratio will stay close to 1. This gives an argument in favour
of an almost spherical dark halo in UGC\,4385 but, within quoted
errors, even flattened halo is consistent with observations.
For instance, assuming a model with a scale-free potential
(see Whitmore et al. 1987), from the observed V$_{pol}$/V$_{eq}$ ratio
one can obtain that 
the flattening of equipotentials in the galaxy is 1.05$\pm$0.27.

\begin{figure}
\centerline{\psfig{file=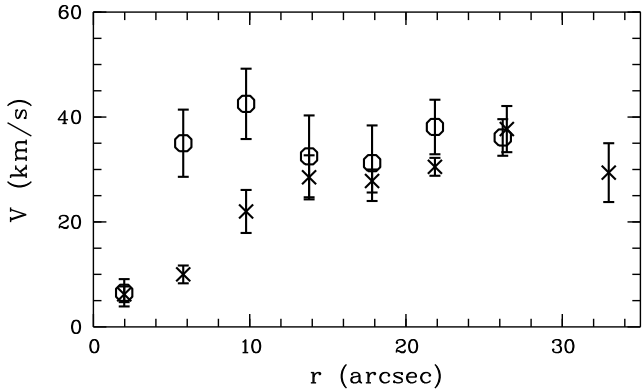,width=8.8cm,angle=-90,clip=}}
\caption{Averaged H$\alpha$ rotation curve of the main body of 
UGC\,4385 at P.A.=17$^{\rm o}$ 
(open circles) and of the polar ring
at P.A.=107$^{\rm o}$ (crosses). Original observations were presented in
Reshetnikov \& Combes (1994). }
\end{figure}

Assuming a spherical mass distribution, the total mass to $B$-luminosity 
ratio within the last point of the rotation curve of the
ring ($r=33\arcsec$) is 0.4$\pm$0.15 (in solar units).

\subsubsection{Main results on UGC\,4385}

Kinematical evidence for two luminous nearly orthogonal components
and the optical morphology suggest that UGC\,4385 is true polar-ring
galaxy. The chaotic appearance of the ring, its protuberant shape,
global assymetry of the
galaxy, strong isophotal twisting all suggest the
relative youth of the object. Comparison with results of numerical
simulations (e.g. Bekki 1997; Bournaud \& Combes 2003) shows that 
in the case of UGC\,4385 we can observe an early stage of PRG 
formation due to galaxy merging. Also, there are no 
comparably sized companions (possible donors for the accretion scenario) 
at least up to 50 kpc from the galaxy.
Most probably, we observe a head-on
collision between two small galaxies. At least one of the galaxies
is a late-type galaxy and we see its perturbed disk as the
polar ring. A red stellar halo is a natural consequence of the merging
process (Bournaud \& Combes 2003). Numerous candidates for young
star clusters (Fig. 13) also support possible ongoing interaction 
of two pre-existing galaxies. 

The maximum rotation velocity of the galaxy derived from the HI
width (Table 2) is $\approx$80 km/s that is in evident contradiction
with optical rotational velocity (V$_{\rm max}\approx$\,40 km/s -- 
Fig. 14). It means, probably, that the HI disk of the galaxy is more
extended than the optical image and that real rotation curve is still
rising beyond the optical disk. As in the cases of A\,0017+2212 and 
UGC\,1198, new kinematical data and detailed numerical
simulations are highly needed for better understanding of this
galaxy. 

\section{Photometric features of PRGs with spiral hosts}

In this section we will discuss available photometric characteristics
for the PRGs with spiral central galaxies. As for ``classical'' PRGs
with early-type gas-free hosts, the main observational features were
summarized by Reshetnikov \& Sotnikova (1997) and by Iodice et al. (2002a,b).

Among relatively regular PRGs with optical rings we have selected three 
galaxies with spiral hosts: NGC\,660 (van Driel et al. 1995, 
Benvenuti et al. 1976), 
ESO\,235-G58 (Buta \& Crocker 1993, Iodice 2001), UGC\,4385. 
Several other objects can be classified as forming rings around
(or inside?) spiral galaxies due to external accretion or
merging: NGC\,3808B, NGC\,6286 (Reshetnikov et al. 1996), 
UGC\,4261 (Reshetnikov et al. 1998), UGC\,5600 (Karataeva et al. 2001),
A\,0017+2212, UGC\,1198 
from the present work. We will consider also NGC\,2748 -- a spiral with 
a very smooth and faint polar ring (Hagen-Thorn et al. 1996).

\subsection{Central galaxies}

The mean $B$-band absolute magnitude of the sample PRGs is between --19\fm0
and --19\fm5 (without or with internal extinction corrections) with a
standard deviation of 1\fm2. Therefore, they are sub-$L^*$ spirals, 
on average.

Fig. 15 shows the observed integrated colors of the host galaxies 
in the two-color diagram. It is evident in the figure that
the optical colors of the spiral hosts
appear very similar to those of normal spiral galaxies. 
After correction for internal extinction (several sample galaxies are
almost edge-on disks) according to the RC3 (de Vaucouleurs et al. 1991) 
prescriptions, we obtain the next average color indices: \\
\hspace*{1.2cm}$\langle B-V \rangle = +0.53 \pm 0.11$ (10 objects),\\
\hspace*{1.2cm}$\langle V-R \rangle = +0.38 \pm 0.05$ (8 objects).\\
Such colors are usual for normal Sbc-Sc spiral galaxies 
(Buta et al. 1994, Buta \& Williams 1995).

In the same figure we present evolutionary track for a model
spiral galaxy computed with the PEGASE.2 code (Fioc \& Rocca-Volmerange 1997).
The model includes the following main ingredients: extended infall
of a zero-metallicity gas (infall timescale is 5 Gyr);
a gas-dependent star formation rate; inclination-averaged extinction
for the disk geometry. The track (dashed line) starts from 1 Gyr age at 
the left side and finishes at 14 Gyr. This model satisfactorily reproduces
the photometric properties of late-type spirals in the local universe
as far as spiral hosts of PRGs. The implied model age of the main
galaxies is $\geq$7 Gyr. The average extinction-corrected colors 
corresponds to 12$\pm$2 Gyr age.

\begin{figure}
\centerline{\psfig{file=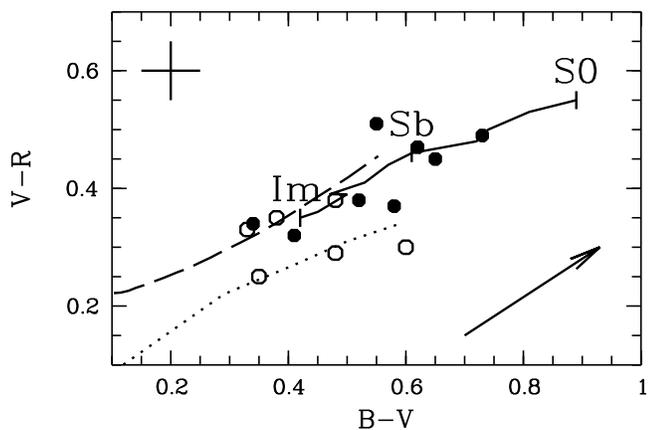,width=8.8cm,angle=-90,clip=}}
\caption{$B-V$ vs. $V-R$ diagram for the main galaxies (solid circles) 
and the polar rings (open circles) of PRGs with spiral hosts. Solid
line represents observational relationship for normal Im--S0 galaxies
according to Buta et al. (1994) and Buta \& Williams (1995).
Dotted line -- simple stellar populations models with different
metallicities at 1 Gyr age (Kurth et al. 1999). Metallicity varies
from $Z=0.0001$ at bottom left end of the line to $Z=0.02$ at top 
right. Dashed line shows evolutionary track for the stellar
synthesis model optimized for a late-type galaxy 
(Fioc \& Rocca-Volmerange 1997). 
Arrow indicates the value and direction of a reddening
for a total $B$-band absorption of 1$^m$, cross demonstrates typical
error of colors measurements.}
\end{figure}

The mean extrapolated central surface brightness of an exponential disk
in the $B$ passband (in this filter the data is available or can be estimated
from published results for all 10 galaxies) is 
$\mu_0(B)=20.2 \pm 0.9$. This is the mean observational value, not corrected
for inclination effects. Adopting the extreme correction for a fully transparent
galaxy (evidently, this is not correct for the $B$ filter), we obtain
the corrected value of $\mu^c_0(B)=21.0 \pm 1.2$. Therefore, stellar 
disks of spiral PRGs are 0\fm7-1\fm5 brighter in comparison with  
normal non-dwarf galaxies (Freeman 1970). The same result was found
earlier for the disks of early-type PRGs (Reshetnikov et al. 1994,
Iodice et al. 2002b). Comparing the PRGs characteristics with 
surveys of local spirals (e.g. O'Neil et al. 2003), we have found that such
bright disks are about 10 times less frequent than ``Freeman'' disks
with $\mu_0(B)=21.7$. 

The probable reason for bright disks in the PRGs
is active star formation triggered by past or current external 
accretion. Other manifestation of recent interaction in spiral
PRGs is the unusual vertical structure of their disks. For instance,
two spiral PRGs with almost edge-on main galaxies (NGC\,3808B and
NGC\,6286) show flaring and very thick (with $h/z_0$ ratio about 2--3)
stellar disks (Reshetnikov et al. 1996).

The average value of the exponential scalelength in the $B$ filter is 
$\langle h \rangle = 1.9 \pm 1.2$ kpc. Spiral galaxies with such
disks are among the most frequent in the local universe
(de Jong 1996).

\subsection{Polar structures}

Typical luminosity of off-plane optical structures is about
0.1--0.2 of the luminosity of spiral hosts. For NGC\,660 and
ESO\,235-G58 the ring-to-host galaxy ratio rises to $\sim$1 in the
$B$-passband.

The average observed colors of polar structures are \\
\hspace*{1.2cm}$\langle B-V \rangle = +0.44 \pm 0.10$ (6 objects), \\
\hspace*{1.2cm}$\langle V-R \rangle = +0.33 \pm 0.05$ (7 objects). \\
Therefore, the polar structures of spiral PRGs, on average, have
bluer colors than the host galaxies but the difference is
not very large (sect. 4.1). As one can see
in Fig. 15, several polar structures demonstrate optical colors
usual for late spirals and for spiral arms of galaxies. In other
cases (NGC\,2748, NGC\,6286, UGC\,4385) the colors are
shifted down from the normal color dependence. The interpretation
of the two-color diagram depends on the adopted metallicity and
star formation history
in the polar structures. For illustrative purposes we present in Fig. 15 
the colors of a simple stellar population model with the age of
1 Gyr (Kurth et al. 1999). The colors of off-plane
structures in NGC\,2748, NGC\,6286, UGC\,4385 are consistent
with this model in the range of metallicities from $Z=0.0004$ to
$Z=0.02$. 

\section{Tully-Fisher relation for PRGs}

The Tully-Fisher (TF) relation is one of the most important
scaling relations for spiral galaxies and it also provides
constraints on galaxy formation. In a recent work, Iodice et al. (2003)
have concluded that the PRG distribution in the TF diagram does
not coincide with the loci occupied by normal disk galaxies. 
(For previous attempts to consider the TF relation for PRGs see
Knapp et al. 1985; Reshetnikov et al. 2001.) Iodice et al. have
concluded that this observational evidence can be explained by a
flattened dark halo, aligned with the polar ring. 
In this secton, we will consider the TF relation using 
a large sample of PRGs (including the new objects presented in this work)
and considering different groups of galaxies separately. 

\subsection{Comparison of the HI widths and velocities}

In order to enlarge the available kinematical data we will combine 
the HI profile widths with rotational velocities measured from
optical emission lines. 
As a first step, we decided to compare the observed 
maximum rotation velocities obtained from emission-line rotation 
curves (V$_{\rm max}$) with the HI global profile widths (W$_{20}$ --
measured HI width at 20\% of peak intensity).  
To make a fair comparison, we have corrected original W$_{20}$
values for the instrumental broadening (following Bottinelli et al. 1990)
and for random motions (according to the prescription of
Tully \& Fouque 1985). For the last correction we have adopted
characteristic parameter values W$_c$=120 km/s and W$_t$=22 km/s
(Rhee 1996, Verheijen 1997).

Fig. 16 shows the comparison of the corrected W$_{20}$ values
(denoted as W$_R$) with observed maximum rotational velocities 
from optical rotation curves. Our sample of PRGs includes 
21 galaxies from the PRC with available HI and optical kinematical 
data. The HI data were taken mainly from van Driel et al. (2000, 2002).
In the case of multiple measurements we preferred the Nacay 
observations. Optical maximum rotation velocities were taken
mainly from the data published in Reshetnikov \& Combes (1994).
For several additional objects the data are from the PRC (A-3, B-19,
B-21), Whitmore et al. (1987) (A-1, A-2, A-5), van Driel et al. (1995)
(C-13), Hagen-Thorn \& Reshetnikov (1997) (B-3), Reshetnikov et al.
(2001) (B-18).

\begin{figure}
\centerline{\psfig{file=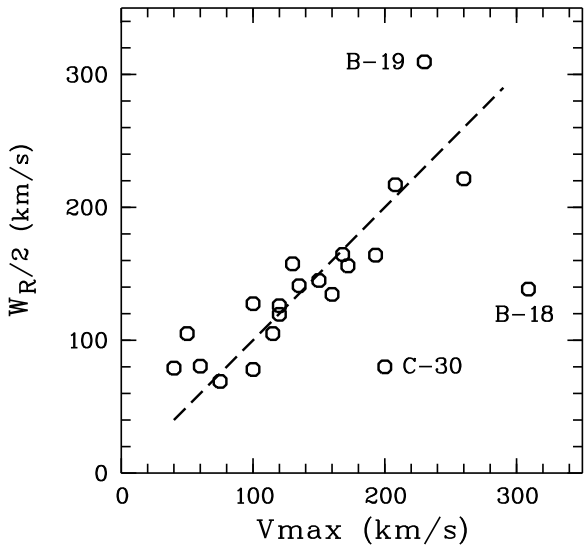,width=8.0cm,angle=-90,clip=}}
\caption{ Comparison of the corrected global HI profile widths with
V$_{\rm max}$ values. Dashed line shows W$_R$/2 = V$_{\rm max}$ relation.}
\end{figure}

After rejecting the three extreme outliers (see Fig. 16), the
mean difference between W$_R$/2 and V$_{\rm max}$ is +2$\pm$25 km/s
(18 galaxies). Excluding dwarf objects with V$_{\rm max}<$\,60 km/s,
we obtain $\langle \frac{W_R}{2}-{\rm V}_{\rm max}\rangle = -4\pm20$ km/s
(16 galaxies). These results show that optical and HI rotation velocities
are well-correlated and therefore internal kinematics measured 
from optical emission lines and from HI data should follow the same
TF relation. 

As for the outliers, in all three cases the HI data are uncertain.
For instance, AM\,1934-563 (B-18) is a member of a triple system
and its HI profile may be confused with another member of the group
(van Driel et al. 2002). The W$_{20}$ value for AM\,2020-504 (B-19) is marked
by van Driel et al. (2002) as uncertain. The UGC\,5101 (C-30) HI profile
(it is in absorption) could be an off-band detection (van Driel et al. 2000).

\subsection{TF relation}

\subsubsection{PRGs}

Fig. 17a shows the Tully--Fisher relation for true polar-ring
galaxies (see description of Category A objects in the PRC). 
We have considered two groups of objects. The first group
consists of galaxies with extended polar rings. This group includes
A\,0136-0801, UGC\,7576, NGC\,4650A, UGC\,9796,
NGC\,660, NGC\,5122, ESO\,503-G17, and ESO\,235-G58. The first six
galaxies are well-known PRGs. ESO\,503-G17 (B-12) is an excellent
candidate PRG whose optical morphology closely resembles
typical PRGs (see PRC). ESO\,235-G58 is also an excellent candidate 
described in detail by Buta \& Crocker (1993) and by Iodice (2001).
The second group consists of objects with relatively short rings:
ESO\,415-G26, NGC\,2685, IC\,1689, AM\,2020-504, ESO\,603-G21, and
UGC\,4385.

The $B$-band magnitudes of the galaxies were taken mainly from
original works with results of detailed surface photometry
(see references in van Driel et al. 2000 and in the present work). 
For A\,0136-0801,
ESO\,503-G17, ESO\,415-G26 and AM\,2020-504 we used the PRC data,
and for NGC\,5122 the magnitude from the NED. The maximum rotation velocities
were estimated from published W$_{20}$ values (see previous section).
For AM\,2020-504 with uncertain HI data (van Driel et al. 2002) 
we used the V$_{\rm max}$ value from the optical rotation curve of the ring
component presented in the PRC. Note that the HI in classic PRGs
associates with the polar ring and, therefore, the observed HI
widths trace the dynamics of the polar structure.

\begin{figure*}
\centerline{\psfig{file=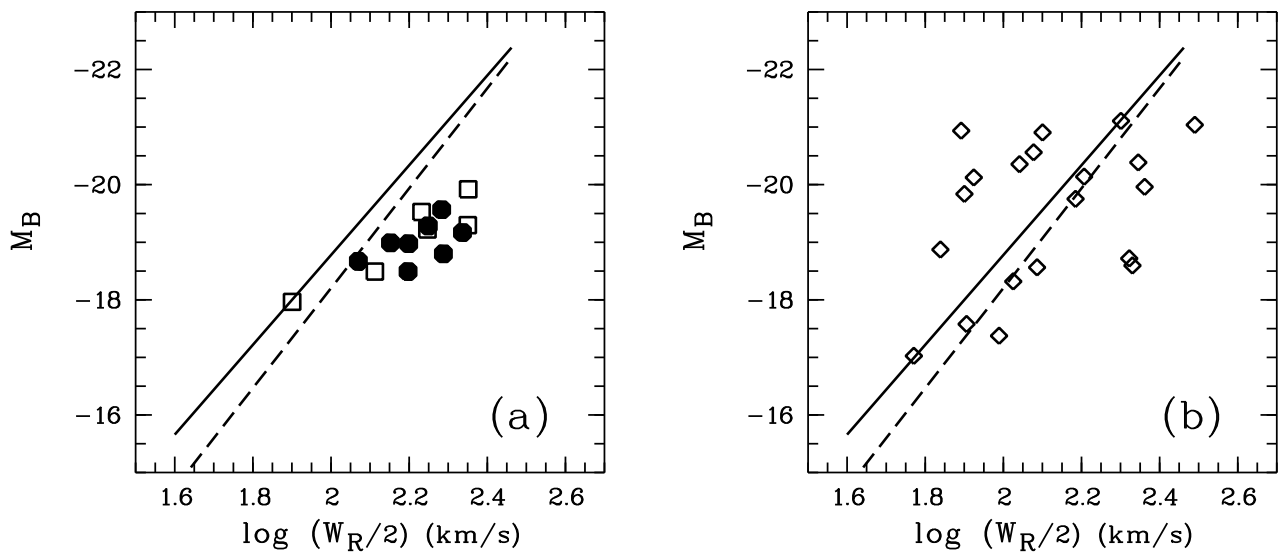,width=17cm,angle=-90,clip=}}
\caption{Absolute $B$-band magnitude vs. the maximum rotation velocity
for PRGs: {\bf a)} for true kinematically-confirmed PRGs (solid 
circles -- galaxies with extended rings, squares -- with short rings);
{\bf b)}  for forming PRGs (rhombs). Solid and dashed lines show
the TF relation for spiral galaxies according to Tully et al. (1998)
and Kannappan et al. (2002) correspondingly (H$_0$=70 km/s Mpc$^{-1}$).}
\end{figure*}

In agreement with Iodice et al. (2003), the PRG characteristics
show a significant shift from the TF relation for normal spirals (Fig. 17a). 
This effect is cleary seen for PRGs with extended and short rings.
Only one galaxy -- UGC\,4385 -- lies on the TF relation. How can possible
internal absorption in the main galaxies (for spiral hosts of NGC\,660 and
ESO\,235-G58) and
in the rings influence the TF relation for PRGs? We have corrected
the total magnitudes for the absorption following Tully et al. (1998).
As a result, two other galaxies -- NGC\,4650A and ESO\,235-G58 -- move
exactly to the TF. The rest of PRGs (11 objects) lie at lower velocities
(for a given luminosity) than those predicted by the TF relation.

Comparing the PRG location with the relations for normal spirals from
Tully et al. (1998) and Kannappan et al. (2002), we have found that
true PRGs demonstrate 35\%$\pm$10\% larger rotation velocities than
disk galaxies. Excluding UGC\,4385, NGC\,4650A, and ESO\,235-G58, we
obtain the mean difference of 45\%$\pm$10\%. 
According to Iodice et al. (2003), these larger velocities can be reproduced
in 2D and 3D N-body models when the dark halo is flattened towards
the polar plane. This result is of primary importance since, if
confirmed, it would imply that some of the dark matter is dissipative,
and set strong constrains on dark matter candidates. 

For galaxies on the TF relation (UGC\,4385, NGC\,4650A, 
ESO\,235-G58), it is interesting to note that the rings in 
all three galaxies are bluer, on average, in comparison with
the ring components in the rest of the objects ($B-V\approx +0.3$ vs.
$B-V\approx$+0.5). One can suppose that these rings are still in the
formation process (see, for instance, sect. 3.3.3).

\subsubsection{Forming PRGs}

Fig. 17b presents the data for 20 objects from the PRC with
available kinematic and photometric data. The objects are:
PRC B-1, 2, 6, 8, 9, 11, 17, 18; C-7, 8, 12, 24, 25, 28, 30, 46,
51; D-12, 18, 19. 
Photometric measurements were taken from original works 
and from the NED. For most of the objects we used the HI measurements
for the rotation velocities estimations (sect. 5.1), while for
AM\,1934-563, UGC\,5101, UGC\,4323, and ESO\,576-G69 we extracted
V$_{\rm max}$ values from optical rotation curves. For evident
spiral galaxies with non-face-on orientation (e.g. NGC\,3808B,
NGC\,6286, AM\,1934-563, NGC\,2748) we have corrected the apparent
magnitudes for internal absorption (Tully et al. 1998).
From the morphological point of view, most of the objects are
interacting galaxies or mergers and so we have called the sample
forming PRGs. 

As it is evident in the figure, forming PRGs follow the standard
TF relation but with rather large scatter. The large scatter can
be explained in part by the very uncertain geometry of the objects.
This leads to very uncertain kinematic and photometric inclination
corrections. Also, in many objects the HI is in the host galaxies and
in possible rings. So we cannot expect that HI widths trace the 
dynamics of the polar structure only.

\section{Conclusions}

In this work we present new $B$, $V$, $R$ photometric data for three 
objects from the PRC (A\,0017+2212, UGC\,1198, UGC\,4385). 
It was shown that all three galaxies are late-type
spiral or irregular galaxies. Main observational peculiarities of the 
objects can be probably explained by current or recent external
accretion or gravitational interaction with a close companion.

Comparison with other PRGs with spiral central galaxies yields that
the host galaxies are, on average, normal sub-$L^*$ Sbc-Sc spirals with 
bright exponential disks. 
The off-plane optical structures have 
somewhat bluer colors than 
the host galaxies.

We have confirmed recent result by Iodice et al. (2003) that 
true evolved polar-ring galaxies with short or extended rings
show larger maximum rotation velocities than expected for the
observed optical luminosity.

\acknowledgements{ I would like to thank the anonymous referee for useful
comments. I acknowledge support from the Russian Foundation 
for Basic Research (03-02-17152) and from the Russian Federal Program 
``Astronomy''
 (40.022.1.1.1101). }



\begin{thebibliography}{}
\bibitem{} Appleton, P.N., \& Struck-Marcell, C. 1996,
Fund. Cosm. Phys., 16, 111
\bibitem{} Bekki, K. 1997, ApJ, 490, L37
\bibitem{} Bekki, K. 1998, ApJ, 499, 635
\bibitem{} Benvenuti, P., Capaccioli, M., \& and D'Odorico, S. 1976,
A\&A, 53, 141
\bibitem{} Borisenko, A.N., Markelov, S.V., \& Ryadchenko, V.P. 1991,
Prepr. SAO $N$ 76
\bibitem{} Bottinelli, L., Gouguenheim, L., Fouque, P., \& Paturel, G.
1990, A\&AS, 82, 391
\bibitem{} Bournaud, F., \& Combes, F. 2003, A\&A, 401, 817
\bibitem{} Bushouse, H.A., Lamb, S.A., \& Werner, M.W. 1988,
ApJ, 335, 74
\bibitem{} Buta, R., \& Crocker, D.A. 1993, AJ, 106, 939
\bibitem{} Buta, R., Mitra, S., de Vaucouleurs, G., Corwin, H.G. 1994,
AJ, 107, 118
\bibitem{} Buta, R., \& Williams, K.L. 1995, AJ, 109, 543
\bibitem{} Buta, R., \& Combes, F. 1996, Fund. Cosm. Phys., 17, 95
\bibitem{} de Jong, T., Clegg, P.E., Soifer, B.T. et al. 1984,
ApJ, 278, L67
\bibitem{} de Jong, R.S. 1996, A\&A, 313, 45
\bibitem{} de Grijs, R. 1998, MNRAS, 299, 595
\bibitem{} de Vaucouleurs, G., de Vaucouleurs, A., Corwin, H.G., et al.
1991, Third Reference Catalogue of Bright Galaxies (Springer-Verlag)
\bibitem{} Fioc, M., \& Rocca-Volmerange, B. 1997, A\&A, 326, 950
\bibitem{} Freeman, K.C. 1970, ApJ, 160, 811
\bibitem{} Galletta, G., Sage, L.J., \& Sparke, L.S. 1997,
MNRAS, 284, 773
\bibitem{} Hagen-Thorn, V.A., Reshetnikov, V.P., \& Yakovleva, V.A. 
1996, Astron. Zh., 73, 36 (Engl. transl. in Astron. Rep.)
\bibitem{} Hagen-Thorn, V.A., \& Reshetnikov, V.P. 1997, A\&A, 319, 430
\bibitem{} Iodice, E. 2001, Ph.D. Thesis, SISSA, Trieste
\bibitem{} Iodice, E., Arnaboldi, M., Sparke, L.S., et al. 2002a,
A\&A, 391, 103
\bibitem{} Iodice, E., Arnaboldi, M., Sparke, L.S., \& Freeman, K.C. 
2002b, A\&A, 391, 117
\bibitem{} Iodice, E., Arnaboldi, M., Bournaud, F. et al. 2003,
ApJ, 585, 730
\bibitem{} Kannappan, Sh.J., Fabricant, D.G., \& Franx, M. 2002,
AJ, 123, 2358
\bibitem{} Karataeva, G.M., Yakovleva, V.A., Hagen-Thorn, V.A. 2001,
Pis'ma v Astron. Zh., 27, 94 (Engl. transl. in Astron. Let.)
\bibitem{} Kinney, A.L., Gallagher, J., Matthews, L. et al. 1999,
AAS, 194.0601
\bibitem{} Knapp, G.R., van Driel W., \& van Woerden H. 1985,
A\&A, 142, 1
\bibitem{} Kurth, O.M., Fritze-v.Alvensleben, U., \& Fricke, K.J. 1999,
A\&AS, 138, 19
\bibitem{} Landolt, A.U. 1983, AJ, 88, 439
\bibitem{} Lonsdale, C.J., Helou, G., Good, J.C. et al. 1989,
Catalogued Galaxies and Quasars in the $IRAS$ Survey, 2nd version,
(JPL, Pasadena)
\bibitem{} O'Neil, K., Andreon, S., \& Cuillandre, J.-C. 2003,
A\&A, 399, L35
\bibitem{} Reshetnikov, V.P., Hagen-Thorn, V.A., \& Yakovleva, V.A.
1993, A\&A, 278, 351
\bibitem{} Reshetnikov, V.P., \& Combes, F. 1994, A\&A, 291, 57
\bibitem{} Reshetnikov, V.P., Hagen-Thorn, V.A., \& Yakovleva, V.A.
1994, A\&A, 290, 693
\bibitem{} Reshetnikov, V.P., Hagen-Thorn, V.A., \& Yakovleva, V.A.
1996, A\&A, 314, 729
\bibitem{} Reshetnikov, V., \& Combes, F. 1997, A\&A, 324, 80
\bibitem{} Reshetnikov, V., \& Sotnikova, N. 1997, A\&A, 325, 933
\bibitem{} Reshetnikov, V.P., Hagen-Thorn, V.A., \& Yakovleva, V.A. 1998,
Astron. Zh., 75, 498 (Engl. transl. in Astron. Rep.)
\bibitem{} Reshetnikov, V.P., Faundez-Abans, M., \& de Oliveira-Abans, M.
2001, MNRAS, 322, 689
\bibitem{} Rhee, M.-H. 1996, PhD thesis, Univ. Groningen
\bibitem{} Schlegel, D.J., Finkbeiner, D.P., Davis, M. 1998,
ApJ, 500, 525
\bibitem{} Schwarzkopf, U., \& Dettmar, R.-J. 2000, A\&A, 361, 451 
\bibitem{} Schweizer, F. 2002, in New Horizonts in Globular Cluster
Astronomy, G.Piotto et al. eds., in press (astro-ph/0212243)
\bibitem{} Skrutskie, M.F., Schneider, S.E., Stiening, R., et al. 1997,
in The Impact of Large Scale Near-IR Sky Surveys, ed. F.Garzon, et al.
(Dordrecht, Kluwer), 25
\bibitem{} Smith, P.S., Januzzi, B.T., \& Elston, R. 1991, ApJS, 77, 67
\bibitem{} Tully, R.B., \& Fouque, P. 1985, ApJS, 58, 67
\bibitem{} Tully, R.B., Pierce, M.J., Huang, J.-Sh., et al. 1998,
AJ, 115, 2264
\bibitem{} van Driel, W., Combes, F., Casoli, F. et al. 1995, AJ, 109, 942
\bibitem{} van Driel, W., Arnaboldi, M., Combes, F., \& Sparke, L.S.
2000, A\&AS, 141, 385
\bibitem{} van Driel, W., Combes, F., Arnaboldi, M., \& Sparke, L.S.
2002, A\&A, 140, 2002
\bibitem{} Verheijen, M.A.W. 1997, PhD thesis, Univ. Gronongen
\bibitem{} Whitmore, B.C., McElroy, D.B., \& Schweizer, F. 1987,
ApJ, 314, 439
\bibitem{} Whitmore, B.C., Lucas, R.A., McElroy, D.B. et al. 1990,
AJ, 100, 1489
 (PRC)

\bibitem{} Young, J.S., \& Knezek, P.M. 1989, ApJ, 347, L55

\end{thebibliography}
\end{document}